\documentstyle[12pt, epsfig]{article}
\newcommand{\bea}   {\begin{eqnarray}}
\newcommand{\eea}   {\end{eqnarray}}
\begin{document}
\renewcommand{\thefootnote}{\fnsymbol{footnote}}

\thispagestyle{empty}

\title{Refining the classification of the irreps of the $1D$ $N$-Extended Supersymmetry}
\author{Zhanna Kuznetsova\thanks{{\em e-mail: zhanna@cbpf.br}} 
~and Francesco Toppan\thanks{{\em e-mail: toppan@cbpf.br}}
\\ \\
{\it $~^\ast$ICE-UFJF, cep 36036-330, Juiz de Fora (MG), Brazil}
\\ 
{\it $~^\dagger$ CBPF, Rua Dr.}
{\it Xavier Sigaud 150,}
 \\ {\it cep 22290-180, Rio de Janeiro (RJ), Brazil}}
\maketitle
\begin{abstract}
The 
linear finite irreducible representations of the algebra of the $1D$ $N$-Extended Supersymmetric Quantum Mechanics are discussed in terms of their ``connectivity" (a symbol encoding information on the graphs associated to the irreps). The classification of the irreducible representations 
with the same fields content and different connectivity is presented up to $N\leq 8$.
\end{abstract}
\vfill 
\rightline{CBPF-NF-020/06}

\newpage
\section{Introduction}

The structure of the irreducible representations of the $N$-extended supersymmetric quantum mechanics has been elucidated 
only recently (see \cite{{pt},{krt},{dfghil},{dfghil2},{top2}}).
One is concerned with the problem of classifying the finite linear irreducible representations
of the supersymmetry algebra
\begin{eqnarray}\label{nsusyalgebra}
\{Q_i,Q_j\}&=& 2\delta_{ij} H,\nonumber\\
\relax [Q_i, H] &=& 0,
\end{eqnarray}
where $Q_i$ are $N$ odd supercharges ($i=1,\ldots ,N$), while the bosonic central extension $H$ can be regarded
as a hamiltonian (therefore $H\equiv i\frac{d}{dt}$) of a supersymmetric quantum mechanical system.
The finite linear irreps of (\ref{nsusyalgebra}) consist of an equal finite number $n$ of bosonic and fermionic
fields (depending on a single coordinate $t$, the time) upon which the supersymmetry operators act linearly.\par
In \cite{pt} it was proven that all (\ref{nsusyalgebra}) irreps fall into classes of equivalence determined by
the irreps of an associated Clifford algebra. As one of the corollaries, a relation between $n$ 
(the total number of bosonic, or fermionic, fields entering the irrep) and the value $N$ of the extended supersymmetry was 
established.\par
A dimensionality $d_i = d_1 + \frac{i-1}{2}$ ($d_1$ is an arbitrary constant) can be assigned to the fields entering an irrep. The difference in dimensionality between a given bosonic and a given fermionic field is a half-integer number. The fields content of an irrep is the set of integers $(n_1,n_2,\ldots , n_l)$ specifying the number $n_i$ of fields of dimension
$d_i$ entering the irrep. Physically, the $n_l$ fields of highest dimension are the auxiliary fields which transform as a time-derivative under any supersymmetry generator. 
The maximal value $l$ (corresponding to the maximal dimensionality
$d_l$) is known as the length of the irrep. 
Either $n_1, n_3,\ldots$ correspond to the bosonic fields (therefore $n_2, n_4, \ldots$ specify the fermionic fields)
or viceversa.  In both cases the equality $n_1+n_3+\ldots =n_2+n_4+\ldots = n$ is guaranteed.
A multiplet is bosonic (fermionic) if its $n_1$ component fields of lower dimensions are bosonic (fermionic).
The representation theory does not discriminate the overall bosonic or fermionic nature of the multiplet. \par
In \cite{krt} the allowed $(n_1,n_2,\ldots, n_l)$ fields contents of the $N$-extended (\ref{nsusyalgebra})
superalgebra were classified
(the results were explicitly furnished for $N\leq 10$). In \cite{top2} it was further pointed out that
an equivalence relation could be introduced in such a way that the fields content uniquely specifies the irreps in
the given class.
On physical grounds, irreps with different fields content produce quite different supersymmetric physical systems.
For instance, the fields content determines the dimensionality of the
target space of the one-dimensional $N$-extended supersymmetric sigma models, see e.g. \cite{abc}. Similarly,
dimensional reductions
of supersymmetric field theories produce extended supersymmetric one-dimensional quantum mechanical systems with 
specific field
contents, see e.g. \cite{top}. 
The classification of the (\ref{nsusyalgebra}) irreps fields contents has very obvious physical meaning. This part of
the program of classifying irreps, due to \cite{krt}, can now be considered completed.\par
The (\ref{nsusyalgebra}) irreps were investigated in \cite{dfghil} in terms of filtered Clifford modules.
In \cite{dfghil} and \cite{dfghil2} it was pointed out that certain irreps admitting the same fields content can be regarded
as inequivalent. These results were obtained by analyzing the ``connectivity properties" (more on that later) of certain graphs
associated to the irreps. A notion of equivalence class among irreps (spotting their difference in 
``connectivity") was introduced. In \cite{dfghil2}, two examples were explicitly presented. They involved a pair
of $N=6$ irreps with $(6,8,2)$ fields content and a pair of $N=5$ irreps with $(6,8,2)$ fields content.
In \cite{dfghil2} the classification of the irreps which differ by connectivity was left as an open problem.\par
In this letter we point out that, using the approach of \cite{krt}, we can easily classify the connectivity 
properties of the irreps of given fields contents. The explicit results will be presented for $N\leq 8$. Since the
$N\leq 4$ cases are trivial, the connectivity being uniquely determined by the fields content, we explicitly present the
results for $N=5,6,7,8$.\par
The connectivity of the irreps (inspired by the graphical presentation of the irreps known as ``Adinkras" \cite{fg})
can be understood as follows. 
For the class of irreducible representations under consideration, any given field of dimension $d$
is mapped, under a supersymmetry transformation, either\par
{\em a}) to a field of dimension $d+\frac{1}{2}$ belonging to the multiplet\footnote{ or to its opposite, the sign of the transformation being irrelevant for our purposes.} or,
\par
{\em b}) to the time-derivative of a field of dimension $d-\frac{1}{2}$.\par
If the given field belongs to an irrep of the $N$-extended (\ref{nsusyalgebra}) supersymmetry algebra, therefore $k\leq N$ of its transformations
are of type {\em a}), while the $N-k$ remaining ones are of type {\em b}). 
Let us now specialize our discussion to a length-$3$ irrep (the interesting case for us). Its fields content is given
by $(n_1,n,n-n_1)$, while the set of its fields is expressed by $(x_i;\psi_j;g_k)$, with $i=1,\ldots , n_1$, $j=1,\ldots ,n$,
$k=1,\ldots , n-n_1$. The $x_i$'s are $0$-dimensional fields (the $\psi_j$ are $\frac{1}{2}$-dimensional and the $g_k$ $1$-dimensional fields,
respectively). 
The connectivity associated to the given multiplet is defined in terms of the $\psi_g$ symbol. It encodes the following
information. The $n$ $\frac{1}{2}$-dimensional fields $\psi_j$ are partitioned in the subsets of $m_r$ fields admitting
$k_r$ supersymmetry transformations of type {\em a}) ($k_r$ can take the $0$ value). We have $\sum_r m_r =n$.
The $\psi_g$ symbol is expressed as
\begin{eqnarray}\label{psig}
\psi_g&\equiv& {m_1}_{k_1}+{m_2}_{k_2}+\ldots
\end{eqnarray}
As an example, the $N=7$ $(6,8,2)$ multiplet admits connectivity $\psi_g= 6_2+2_1$ (see 
(\ref{conn})). It means that there are two types of fields $\psi_j$. $6$ of them are mapped,
under supersymmetry transformations, in the two auxiliary fields $g_k$. The two remaining
fields $\psi_j$ are only mapped into a single auxiliary field.\par  
An analogous symbol, $x_\psi$,
can be introduced. It describes the supersymmetry transformations of the $x_i$ fields into
the $\psi_j$ fields. This symbol is, however, always trivial. An $N$-irrep with $(n_1,n, n-n_1)$ fields content
always produce $x_\psi \equiv {n_1}_N $.
\par
Using the methods of \cite{krt}, we are able to classify here the admissible $\psi_g$ connectivities of the irreps.
The pair of $N=6$ $(6,8,2)$ irreps and the pair of $N=5$ $(6,8,2)$ irreps of \cite{dfghil2} fall into the two admissible
classes of $\psi_g$ connectivity for the corresponding values of $N$ and fields content. \par
In \cite{dfghil2} the two sets of
three ordered numbers (for length-$3$ multiplets), $S=[s_1,s_2,s_3]$ and $T=[t_1,t_2,t_3]$, the ``sources" and ``targets"
respectively, have been introduced. The integer $s_i$ gives the number of fields of dimension $d_i=\frac{i-1}{2}$ which do
not result as an\\ {\em a })-supersymmetry transformation of at least one field of dimension $d_i-\frac{1}{2}$.  
The integer $t_i$ gives the number of fields of dimension $d_i=\frac{i-1}{2}$ which only admit 
supersymmetry transformations of type {\em b}). For a multiplet of $(n_1,n,n-n_1)$ fields content,
necessarily $s_1=n_1$, $s_3=0$, together with $t_1=0$ and $t_3=n-n_1$. $S,$ and $T$ are fully determined once
$s_2$ and $t_2$, respectively, are known.
The complete list of $\psi_g$ connectivities for length-$3$ multiplets contains more information than $S$ and $T$. As for the targets, it is obvious that $t_2$ can be recovered from $\psi_g$. As for the sources, using the $(n_1,n,n-n_1)\leftrightarrow
(n-n_1,n,n_1)$ irreps duality discussed in \cite{krt}, $s_2$ is recovered from the $\psi_g$ connectivity of the
associated dual multiplet. 
In Section {\bf 3} we produce the list of the allowed connectivities.
We prove that the connectivity symbol $\psi_g$ allows to discriminate inequivalent irreps which are not discriminated by the
sources and targets $S$, $T$ (of given heighth) introduced in \cite{dfghil2}. 
In Section {\bf 4} we 
summarize the previous results, presenting the full list of $N\leq 8$ irreps differing by sources and targets,
as well as the full list of $N\leq 8$ irreps with the same sources and targets and different $\psi_g$ connectivity. We 
explicitly 
present the $N=5$ supersymmetry transformations for two such irreps. We also present them graphically (the associated ``adinkras").
We postpone to the Conclusions a discussion of the possible interpretations of our finding.\par
This paper is structured as follows. In the next Section, the needed ingredients and \cite{krt} conventions are reviewed.
The main results are presented in Section {\bf 3}. The irreps connectivities are furnished for all cases which can
potentially produce inequivalent results (therefore, for the  
$N=5,6,7$ length-$3$ and length-$4$ irreps). In Section {\bf 4} it is pointed out that the $\psi_g$ connectivities computed
in Section {\bf 3} can discriminate irreps which are not discriminated by the sets of ``sources and targets" numbers
employed in \cite{dfghil2}. Further comments and open problems are discussed in the Conclusions. To make the paper
self-consistent, an Appendix with our conventions of the $Cl(0,7)$ Clifford generators (used to construct the 
$N=5,6,7,8$ supersymmetry operators) is added.

\section{Basic notions and conventions}

In this Section we summarize the basic notions, results and conventions of \cite{krt} that will be needed in the following.
Up to $N\leq 8$, inequivalent connectivities are excluded for $N=1,2,3,4$ and can only appear, in principle, for $N=5,6,7,8$.
The irreps of the $N=5,6,7,8$ supersymmetric extensions can be obtained through a dressing of the $N=8$ length-$2$ root multiplet (see \cite{krt} and the comment in \cite{top2}). For simplicity, we can therefore limit the discussion of the \cite{krt} construction starting from the $N=8$
length-$2$ root multiplet. It involves $8$ bosonic and $8$ fermionic fields entering a column vector (the bosonic fields are accommodated in the upper part). The $8$ supersymmetry
operators ${\widehat Q}_i$ ($i=1,\ldots ,8)$ in the $(8,8)$ $N=8$ irrep are given by the matrices
\begin{eqnarray}\label{hatq}
{\widehat Q}_j = \left(
\begin{array}{cc}
0&\gamma_j\\
-\gamma_j\cdot H&0
\end{array}
\right), &&
{\widehat Q}_8 = \left(
\begin{array}{cc}
0&{\bf 1}_8\\
{\bf 1}_8\cdot H&0
\end{array}
\right)
\end{eqnarray}
where  the $\gamma_j$ matrices ($j=1,\ldots , 7$) are the $8\times 8$ generators of the $Cl(0,7)$ Clifford algebra
and $H=i\frac{d}{dt}$ is the hamiltonian.  The $Cl(0,7)$ Clifford irrep is uniquely defined up to similarity transformations
and an overall sign flipping \cite{oku}. Without loss of generality we can unambiguously fix the $\gamma_j$ matrices to be given
as in the Appendix. Each $\gamma_j$ matrix (and the ${\bf 1}_8$ identity) possesses $8$ non-vanishing entries, one in each column and one in each row. The whole set of non-vanishing entries of the eight (\ref{gammas07}) matrices
fills the entire $8\times 8=64$ squares of a ``chessboard". The chessboard appears in the upper right block of (\ref{hatq}).\par
The length-$3$ and length-$4$ $N=5,6,7,8$ irreps (no irrep with length $l>4$ exists for $N\leq 9$, see \cite{krt}) are acted upon by the $Q_i$'s supersymmetry transformations, obtained from the original ${\widehat Q}_i$ operators through a dressing,
\begin{eqnarray}
{\widehat Q}_i &\rightarrow& Q_i = D{\widehat Q}_i D^{-1},
\end{eqnarray} 
realized by a diagonal dressing matrix $D$. It should be noticed that only the subset of ``regular" dressed operators $Q_i$ (i.e., having no $\frac{1}{H}$ or higher poles in its entries) act on the new irreducible multiplet. 
Apart from the self-dual $(4,8,4)$ $N=5,6$ irreps, 
without loss of generality, for our purpose of computing the irreps connectivities, the diagonal dressing matrix
$D$ which produces an irrep with $(n_1,n,n-n_1)$ fields content can be chosen to have its non-vanishing diagonal entries
given by $\delta_{pq} d_q$, with $d_q =1$ for $q=1,\ldots, n_1$ and $q=n+1,\ldots , 2n$, while $d_q=H$ for $q=n_1+1,\ldots, n$. Any permutation of the first $n$ entries produces a dressing which is equivalent,
for computing both
the fields content and the $\psi_g$ connectivity, to $D$.
The only exceptions correspond to the $N=5$ $(4,8,4)$ and $N=6$ $(4,8,4)$ irreps. Besides the diagonal
matrix $D$ as above, inequivalent irreps can be obtained by a diagonal dressing $D'$ with diagonal
entries $\delta_{pq}{d'}_q$, with ${d'}_q=H$ for $q=4,6,7,8$ and ${d'}_q=1$ for the remaining values of $q$.
\par
Similarly, the $(n_1, n_2, n-n_1, n-n_2)$ length-$4$ multiplets are acted upon by the $Q_i$ operators dressed by $D$, whose
non-vanishing diagonal entries are now given by $\delta_{pq}d_q$, with 
$d_q=1$ for $q=1,\ldots, n_1$ and $q=2n-n_2+1,\ldots, 2n$, while $d_q=H$ for $q=n_1+1,\ldots, 2n-n_2$.  
\par
The $N=5,6,7,8$ length-$2$ $(8,8)$ irreps are unique (for the given value of $N$), see \cite{top2}.\par
It is also easily recognized that all $N=8$ length-$3$ irreps of given fields content produce the same value
of $\psi_g$ connectivity (\ref{psig}).  For what concerns the length-$3$ $N=5,6,7$ irreps the situation is as follows.
Let us consider the irreps with $(k, 8, 8-k)$ fields content. Its supersymmetry transformations are defined by picking
an $N<8$ subset from the complete set of $8$ dressed $Q_i$ operators.
It is easily recognized that for $N=7$, no matter which supersymmetry operator is discarded, any choice of the
seven operators produces the same value for the $\psi_g$ connectivity. Irreps with different connectivity can therefore
only be found for $N=5,6$. The $\left(\begin{array}{c}8\\6\end{array}\right)=28$ choices of $N=6$ operators fall into two
classes, denoted as $A$ and $B$, which can, potentially, produce $(k, 8, 8-k)$ irreps with different connectivity. Similarly, the  $\left(\begin{array}{c}8\\5\end{array}\right)=56$ choices of $N=5$ operators 
fall into two $A$ and $B$ classes which can, potentially, produce irreps of different connectivity. For some given $(k,8,8-k)$ irrep,
the value of $\psi_g$ connectivity computed in both $N=5$ (as well as $N=6$) classes
can actually coincide. In the next Section we will show when this feature indeed happens.\par
To be specific, we present a list of representatives of the supersymmetry operators for each $N$ and in each 
$N=5,6$ $A,B$ class. We have, with diagonal dressing $D$,
\begin{eqnarray}\label{susyop}
&\begin{array}{cll}
N=8 &\equiv& Q_1,Q_2,Q_3,Q_4,Q_5,Q_6,Q_7,Q_8\\
N=7 &\equiv& Q_1,Q_2,Q_3,Q_4,Q_5,Q_6,Q_7\\
N=6 ~ (case~ A)&\equiv& Q_1,Q_3,Q_4,Q_5,Q_6,Q_7\\
N=6 ~ (case ~B)&\equiv& Q_1, Q_2,Q_3,Q_4,Q_5,Q_6\\
N=5 ~ (case ~A) &\equiv& Q_3,Q_4,Q_5,Q_6,Q_7\\
N=5 ~ (case ~B) &\equiv& Q_2,Q_3,Q_4,Q_5,Q_6
\end{array}
&\end{eqnarray}
and, with diagonal dressing $D'$ for the $(4,8,4)$ irreps,
\begin{eqnarray}\label{susyop2}
&\begin{array}{cll}
N=6 ~ (case~ A')&\equiv& Q_1,Q_3,Q_4,Q_5,Q_6,Q_7\\
N=5 ~ (case ~A') &\equiv& Q_3,Q_4,Q_5,Q_6,Q_7
\end{array}
&\end{eqnarray}
We are now in the position to compute the connectivities of the irreps (the results are furnished in the next Section).
Quite literally, the computations can be performed by filling a chessboard with pawns representing the allowed configurations.
\section{Classification of the irreps connectivities}

In this Section we report the results of the computation of the allowed connectivities for the $N=5,6,7$ 
length-$3$ and length-$4$ irreps. As discussed in the previous Section, the only values of $N\leq 8$ which allow
the existence of multiplets with the same fields content but inequivalent connectivities are $N=5$ and $N=6$.
We also produce the $S$ and $T$ allowed sources and targets numbers for the irreps. As recalled in the Introduction, 
the $S$ sources can be recovered from a symbol, denoted as ``$~_x\psi$", expressing the partitions of the $n$ 
$\frac{1}{2}$-dimensional fields $\psi_j$ in terms of the $h_r\leq N$ number of supersymmetry transformations
of {\em a}) type which map the $x_i$ fields on a given $\frac{1}{2}$-dimensional field. Due to the irrep
$(n_1,n,n-n_1)\leftrightarrow (n-n_1,n,n_1)$  duality discussed in \cite{krt}, $~_x\psi$ is recovered from the
$\psi_g$ connectivity of its dual irrep. Indeed 
\begin{eqnarray}
~_x\psi [(k,n,n-k)_\ast] &=&\psi_g [(n-k,n,k)_\ast]
\end{eqnarray}
(the suffix $\ast \equiv A, B$ has been introduced in order to discriminate, when needed, the $A$ and $B$ subcases of $N=5,6$).
\par
Our results concerning the allowed $\psi_g$ connectivities of the length-$3$ irreps are reported in the following table
(the $A, A', B$ cases of $N=5,6$ are specified)
\begin{eqnarray}\label{conn}
&\begin{array}{|c|c|l|l|}\hline
length-3&{N=7}&{N=6}&{N=5}\\ \hline 
(7,8,1)&7_1+1_0& 6_1+2_0&5_1+3_0\\ \hline
(6,8,2)&6_2+2_1& 6_2+2_0~{(A)}&4_2+2_1+2_0~{(A)}\\ 
 && 4_2+4_1~{(B)}&2_2+6_1~{(B)}\\ \hline
(5,8,3)&5_3+3_2&4_3+2_2+2_1~{(A)}&4_3+3_1+1_0~{(A)}\\ 
&&2_3+6_2~{(B)}&1_3+5_2+2_1~{(B)}\\ \hline
(4,8,4)&4_4+4_3&4_4+4_2~{(A)}&4_4+4_1~{(A)}\\ 
&&2_4+4_3+2_2~{(A')}&1_4+3_3+3_2+1_1~{(A')}\\ 
&&8_3~{(B)}&4_3+4_2~{(B)}\\ \hline
(3,8,5)&3_5+5_4&2_5+2_4+4_3~{(A)}&1_5+3_4+4_2~{(A)}\\ 
&&6_4+2_3~{(B)}&2_4+5_3+1_2~{(B)}\\ \hline
(2,8,6)&2_6+6_5&2_6+6_4~{(A)}&2_5+2_4+4_3~{(A)}\\ 
&&4_5+4_4~{(B)}&6_4+2_3~{(B)}\\ \hline
(1,8,7)&1_7+7_6&2_6+6_5&3_5+5_4\\ \hline
 \end{array}
 &\nonumber\\
 &&
 \end{eqnarray}

The $\psi_g$ connectivities of the $N=5$ (and $N=6$)  $A$ and $B$ subcases collapse to the same value
for the $(1,8,7)$ and $(7,8,1)$ irreps, proving that these multiplets do not admit inequivalent connectivities. \par
It is helpful to produce tables with the values of the $\psi_g$ connectivity, the $S$ sources and the $T$ targets for the irreps
admitting inequivalent connectivities. For $N=6$ we get
\begin{eqnarray}\label{connN6}
&\begin{array}{|c|c|c|c|}\hline
N=6:&connectivities&sources&targets\\ \hline
(6,8,2)_A& 6_2+2_0& S=[6,0,0] &T =[0,2,2]\\
(6,8,2)_B&4_2+4_1& S=[6,0,0]&T=[0,0,2]\\ \hline
(5,8,3)_A& 4_3+2_2+2_1& S=[5,0,0]& T =[0,0,3]\\
(5,8,3)_B&2_3+6_2& S=[5,0,0]&T=[0,0,3] \\ \hline
(4,8,4)_A& 4_4+4_2& S=[4,0,0] &T =[0,0,4]\\
(4,8,4)_{A'}& 2_4+4_3+2_2& S=[4,0,0] &T =[0,0,4]\\
(4,8,4)_B&8_3& S=[4,0,0]&T=[0,0,4]\\ \hline
(3,8,5)_A& 2_5+2_4+4_3& S=[3,0,0]& T =[0,0,5]\\
(3,8,5)_B&6_4+2_3& S=[3,0,0]&T=[0,0,5] \\ \hline
(2,8,6)_A& 2_6+6_4& S=[2,2,0]& T =[0,0,6]\\
(2,8,6)_B&4_5+4_4& S=[2,0,0]&T=[0,0,6] \\ \hline
\end{array}
&
\end{eqnarray}
For $N=5$ we obtain
\begin{eqnarray}\label{connN5}
&\begin{array}{|c|c|c|c|}\hline
N=5:&connectivities&sources&targets\\ \hline
(6,8,2)_A& 4_2+2_1+2_0& S=[6,0,0] &T =[0,2,2]\\
(6,8,2)_B&2_2+6_1& S=[6,0,0]&T=[0,0,2]\\ \hline
(5,8,3)_A& 4_3+3_1+1_0& S=[5,0,0]& T =[0,1,3]\\
(5,8,3)_B&1_3+5_2+2_1& S=[5,0,0]&T=[0,0,3] \\ \hline
(4,8,4)_A& 4_4+4_1& S=[4,0,0] &T =[0,0,4]\\
(4,8,4)_{A'}& 1_4+3_3+3_2+1_1& S=[4,0,0] &T =[0,0,4]\\
(4,8,4)_B&4_3+4_2& S=[4,0,0]&T=[0,0,4]\\ \hline
(3,8,5)_A& 1_5+3_4+4_2& S=[3,1,0]& T =[0,0,5]\\
(3,8,5)_B&2_4+5_3+1_2& S=[3,0,0]&T=[0,0,5] \\ \hline
(2,8,6)_A& 2_5+2_4+4_3& S=[2,2,0]& T =[0,0,6]\\
(2,8,6)_B&6_4+2_3& S=[2,0,0]&T=[0,0,6] \\ \hline
\end{array}
&
\end{eqnarray}
We postpone to Section {\bf 4} the discussion of our results.

\subsection{Connectivities of the length-$4$ multiplets}

Up to $N\leq 8$, the only admissible $(n_1,n_2,n-n_1,n-n_2)$ length-$4$ fields contents for the
$(x_i; \psi_j; g_k;\omega_l)$ irreps are given below (see \cite{krt}). 
Here $x_i$ ($i=1,\ldots , n_1$) denote the $0$-dimensional fields, $\psi_j$ ($j=1, \ldots , n_2$)
denote the $\frac{1}{2}$-dimensional fields, $g_k$ ($k=1,\ldots , n-n_1$) denote the $1$-dimensional
fields and, finally, $\omega_l$ ($l=1,\ldots n-n_2$) denote the $\frac{3}{2}$-dimensional auxiliary fields.\par
The analysis of
the connectivities of the length-$4$ irreps is done as in the case of the length-$3$ irreducible multiplets. 
Since we have an extra set of fields w.r.t. the length-$3$ multiplets, the results can be expressed
in terms of one more non-trivial symbol. Besides $\psi_g$, we introduce the $g_\omega$ symbol as well.
The definition of $g_\omega$ follows the definition of $\psi_g$ in (\ref{psig}). The difference of
$g_\omega$ w.r.t. $\psi_g$ is that the $g_k$ fields enter
now in the place of the $\psi_j$ fields, while the $\omega_l$ fields enter in the place of the $g_k$ fields.\par  
Contrary to the case of the length-$3$ irreps, the connectivity of the length-$4$ irreps is uniquely specified in terms of
$N$ and the length-$4$ fields content. The complete list of results is presented in the following table
\begin{eqnarray}
&\begin{array}{|cc|c|c|}\hline
length-4& su.sies& \psi_g&g_\omega\\ \hline 
(1,7,7,1):& N=7&7_6&7_1\\
& N=6&1_6+6_5&6_1+1_0\\ 
& N=5&2_5+5_4&5_1+2_0\\ \hline
(2,7,6,1):
& N=6&1_6+6_4&6_1\\ 
& N=5&1_5+2_4+4_3&5_1+1_0\\ \hline
(2,6,6,2):
& N=6&6_4&6_2\\ 
& N=5&2_4+4_3&4_2+2_1\\ \hline
(1,6,7,2):
& N=6&6_5&6_2+1_0\\ 
& N=5& 1_5+5_4&4_2+2_1+1_0\\ \hline
(1,5,7,3):&
N=5& 5_4&4_3+3_1\\ \hline
(3,7,5,1):
&N=5& 3_4+4_2&5_1\\ \hline
(1,3,3,1):
&N=3& 3_2&3_1\\ \hline
\end{array}
&
\end{eqnarray}

\section{On ``irreps connectivities" versus ``sources and targets"}

From the results presented in (\ref{connN6}) and (\ref{connN5}) we obtain two corollaries.
At first we notice that, besides the $N=6$ $(6,8,2)$ and $N=5$ $(6,8,2)$ pairs of cases presented in \cite{dfghil2}, there exists
four extra pairs, for $N\leq 8$, of inequivalent irreps with the same fields content which differ by the values
of the sources and targets. The whole list of such pairs is given by
\begin{eqnarray}\label{ST}
N=6: &(6,8,2)_A\leftrightarrow (6,8,2)_B&\nonumber\\
N=6: &(2,8,6)_A\leftrightarrow (2,8,6)_B&\nonumber\\
N=5: &(6,8,2)_A\leftrightarrow (6,8,2)_B&\nonumber\\
N=5: &(5,8,3)_A\leftrightarrow (5,8,3)_B&\nonumber\\
N=5: &(3,8,5)_A\leftrightarrow (3,8,5)_B&\nonumber\\
N=5: &(2,8,6)_A\leftrightarrow (2,8,6)_B&
\end{eqnarray}
The above list produces the complete classification of inequivalent $N\leq 8$ irreps that
are discriminated by different values of $S$ and $T$ {\em alone}.    \par

On the other hand, a second corollary of the (\ref{connN6}) and (\ref{connN5}) results shows the existence of extra
irreps sharing the same fields content $(n_1,n,n-n_1)$, the same sources $S=[s_1,s_2,s_3]$ and 
the same targets $T=[t_1,t_2,t_3]$
which, nevertheless, admit different $\psi_g$ connectivity. They are given by
\begin{eqnarray}\label{connST}
N=6: &(3,8,5)_A\leftrightarrow (3,8,5)_B&\nonumber\\
N=6: &(4,8,4)_A\leftrightarrow (4,8,4)_{A'}\leftrightarrow(4,8,4)_B&\nonumber\\
N=6: &(5,8,3)_A\leftrightarrow (5,8,3)_B&\nonumber\\
N=5: &(4,8,4)_A\leftrightarrow (4,8,4)_{A'}\leftrightarrow(4,8,4)_B&
\end{eqnarray}

In order to convince the reader of the existence of such irreps with same sources and targets
but different connectivity
it is useful to explicitly present the supersymmetry transformations 
(depending on the $\varepsilon_i$ global fermionic parameters) in at least one case. We write below a pair of $N=5$ irreps
(the $(4,8,4)_A$ and the $(4,8,4)_B$ multiplets) differing
by connectivity, while admitting the same number of sources and the same number of targets. 
It is also convenient to visualize them graphically as adinkras (see \cite{fg}). The graphical presentation at the end of this Section is given as follows.
Three rows of (from bottom to up) $4$, $8$ and $4$ dots are associated with the $x_i$, $\psi_j$ and $g_k$ fields, respectively. Supersymmetry transformations are represented by lines of $5$ different colors (since $N=5$). Solid lines are associated to transformations with a positive
sign, dashed lines with a negative sign. It is easily recognized that in the type $A$ graph
there are $4$ $\psi_j$ points with four colored lines connecting them to the $g_k$ points,
while the $4$ remaining $\psi_j$ points admit a single line connecting them to the $g_k$ points.
In the type $B$ graph we have $4$ $\psi_j$ points with three colored lines and the $4$
remaining $\psi_j$ points with two colored lines connecting them to the $g_k$ points.\par
The supersymmetry transformations are explicitly given by
\newpage
{\em i) The $N=5$ $(4,8,4)_A$ transformations:}
\begin{eqnarray}\label{484A}
\delta{x_1}&=& \varepsilon_2\psi_3+\varepsilon_4\psi_5+\varepsilon_3\psi_6+\varepsilon_1\psi_7+\varepsilon_5\psi_8\nonumber\\
\delta{x_2}&=& \varepsilon_2\psi_4+\varepsilon_3\psi_5-\varepsilon_4\psi_6-\varepsilon_5\psi_7+\varepsilon_1\psi_8\nonumber\\
\delta{x_3}&=&-\varepsilon_2\psi_1-\varepsilon_1\psi_5-\varepsilon_5\psi_6+\varepsilon_4\psi_7+\varepsilon_3\psi_8 \nonumber\\
\delta{x_4}&=&-\varepsilon_2\psi_2+\varepsilon_5\psi_5-\varepsilon_1\psi_6+\varepsilon_3\psi_7-\varepsilon_4\psi_8 \nonumber\\
\delta{\psi_1}&=&
-i\varepsilon_2{\dot x}_3-\varepsilon_4 g_1-\varepsilon_3g_2-\varepsilon_1g_3-\varepsilon_5g_4
\nonumber\\
\delta{\psi_2}&=& 
-i\varepsilon_2{\dot x}_4-\varepsilon_3 g_1+\varepsilon_4g_2+\varepsilon_5g_3-\varepsilon_1g_4\nonumber\\
\delta{\psi_3}&=& 
i\varepsilon_2{\dot x}_1+\varepsilon_1 g_1+\varepsilon_5g_2-\varepsilon_4g_3-\varepsilon_3g_4\nonumber\\
\delta{\psi_4}&=&
i\varepsilon_2{\dot x}_2-\varepsilon_5 g_1+\varepsilon_1g_2-\varepsilon_3g_3+\varepsilon_4g_4 \nonumber\\
\delta{\psi_5}&=& 
i\varepsilon_4{\dot x}_1+i\varepsilon_3 {\dot x}_2-i\varepsilon_1{\dot x}_3+i\varepsilon_5{\dot x}_4
+\varepsilon_2g_3\nonumber\\
\delta{\psi_6}&=& 
i\varepsilon_3{\dot x}_1-i\varepsilon_4 {\dot x}_2-i\varepsilon_5{\dot x}_3-i\varepsilon_1{\dot x}_4
+\varepsilon_2g_4 \nonumber\\
\delta{\psi_7}&=&  
i\varepsilon_1{\dot x}_1-i\varepsilon_5 {\dot x}_2+i\varepsilon_4{\dot x}_3+i\varepsilon_3{\dot x}_4
-\varepsilon_2g_1\nonumber\\
\delta{\psi_8}&=& 
i\varepsilon_5{\dot x}_1+i\varepsilon_1 {\dot x}_2+i\varepsilon_3{\dot x}_3-i\varepsilon_4{\dot x}_4
-\varepsilon_2g_2 \nonumber\\
\delta{g_1}&=&-i\varepsilon_4{\dot \psi}_1-i\varepsilon_3{\dot\psi}_2+i\varepsilon_1{\dot\psi}_3-i\varepsilon_5{\dot\psi}_4-i\varepsilon_2{\dot\psi}_7  \nonumber\\
\delta{g_2}&=& -i\varepsilon_3{\dot \psi}_1+i\varepsilon_4{\dot\psi}_2+i\varepsilon_5{\dot\psi}_3+i\varepsilon_1{\dot\psi}_4-i\varepsilon_2{\dot\psi}_8 \nonumber\\
\delta{g_3}&=& -i\varepsilon_1{\dot \psi}_1+i\varepsilon_5{\dot\psi}_2-i\varepsilon_4{\dot\psi}_3-i\varepsilon_3{\dot\psi}_4+i\varepsilon_2{\dot\psi}_5 \nonumber\\
\delta{g_4}&=& -i\varepsilon_5{\dot \psi}_1-i\varepsilon_1{\dot\psi}_2-i\varepsilon_3{\dot\psi}_3+i\varepsilon_4{\dot\psi}_4+i\varepsilon_2{\dot\psi}_6 
\end{eqnarray}

{\em ii) The $N=5$ $(4,8,4)_B$ transformations:}

\begin{eqnarray}\label{484B}
\delta{x_1}&=& \varepsilon_5\psi_2+\varepsilon_2\psi_3+\varepsilon_4\psi_5+\varepsilon_3\psi_6+\varepsilon_1\psi_7\nonumber\\
\delta{x_2}&=&-\varepsilon_5\psi_1+ \varepsilon_2\psi_4+\varepsilon_3\psi_5-\varepsilon_4\psi_6+\varepsilon_1\psi_8\nonumber\\
\delta{x_3}&=&-\varepsilon_2\psi_1-\varepsilon_5\psi_4-\varepsilon_1\psi_5+\varepsilon_4\psi_7+\varepsilon_3\psi_8 \nonumber\\
\delta{x_4}&=&-\varepsilon_2\psi_2+\varepsilon_5\psi_3-\varepsilon_1\psi_6+\varepsilon_3\psi_7-\varepsilon_4\psi_8 \nonumber\\
\delta{\psi_1}&=&-i\varepsilon_5{\dot x}_2
-i\varepsilon_2{\dot x}_3-\varepsilon_4 g_1-\varepsilon_3g_2-\varepsilon_1g_3
\nonumber\\
\delta{\psi_2}&=&i\varepsilon_5{\dot x}_1 
-i\varepsilon_2{\dot x}_4-\varepsilon_3 g_1+\varepsilon_4g_2-\varepsilon_1g_4\nonumber\\
\delta{\psi_3}&=& 
i\varepsilon_2{\dot x}_1+i\varepsilon_5{\dot x}_4+\varepsilon_1 g_1-\varepsilon_4g_3-\varepsilon_3g_4\nonumber\\
\delta{\psi_4}&=&
i\varepsilon_2{\dot x}_2-i\varepsilon_5{\dot x}_3+\varepsilon_1g_2-\varepsilon_3g_3+\varepsilon_4g_4 \nonumber\\
\delta{\psi_5}&=& 
i\varepsilon_4{\dot x}_1+i\varepsilon_3 {\dot x}_2-i\varepsilon_1{\dot x}_3-\varepsilon_5g_2
+\varepsilon_2g_3\nonumber\\
\delta{\psi_6}&=& 
i\varepsilon_3{\dot x}_1-i\varepsilon_4 {\dot x}_2-i\varepsilon_1{\dot x}_4
+\varepsilon_5g_1
+\varepsilon_2g_4 \nonumber\\
\delta{\psi_7}&=&  
i\varepsilon_1{\dot x}_1+i\varepsilon_4{\dot x}_3+i\varepsilon_3{\dot x}_4
-\varepsilon_2g_1+\varepsilon_5g_4\nonumber\\
\delta{\psi_8}&=& i\varepsilon_1 {\dot x}_2+i\varepsilon_3{\dot x}_3-i\varepsilon_4{\dot x}_4
-\varepsilon_2g_2 -\varepsilon_5g_3\nonumber\\
\delta{g_1}&=&-i\varepsilon_4{\dot \psi}_1-i\varepsilon_3{\dot\psi}_2+i\varepsilon_1{\dot\psi}_3+i\varepsilon_5{\dot\psi}_6-i\varepsilon_2{\dot\psi}_7  \nonumber\\
\delta{g_2}&=& -i\varepsilon_3{\dot \psi}_1+i\varepsilon_4{\dot\psi}_2+i\varepsilon_1{\dot\psi}_4-i\varepsilon_5{\dot\psi}_5-i\varepsilon_2{\dot\psi}_8 \nonumber\\
\delta{g_3}&=& -i\varepsilon_1{\dot \psi}_1-i\varepsilon_4{\dot\psi}_3-i\varepsilon_3{\dot\psi}_4+i\varepsilon_2{\dot\psi}_5-i\varepsilon_5{\dot\psi}_8 \nonumber\\
\delta{g_4}&=& -i\varepsilon_1{\dot\psi}_2-i\varepsilon_3{\dot\psi}_3+i\varepsilon_4{\dot\psi}_4+i\varepsilon_2{\dot\psi}_6
+i\varepsilon_5{\dot \psi}_7 
\end{eqnarray}
\begin{figure}[htbp]
\epsfig{file=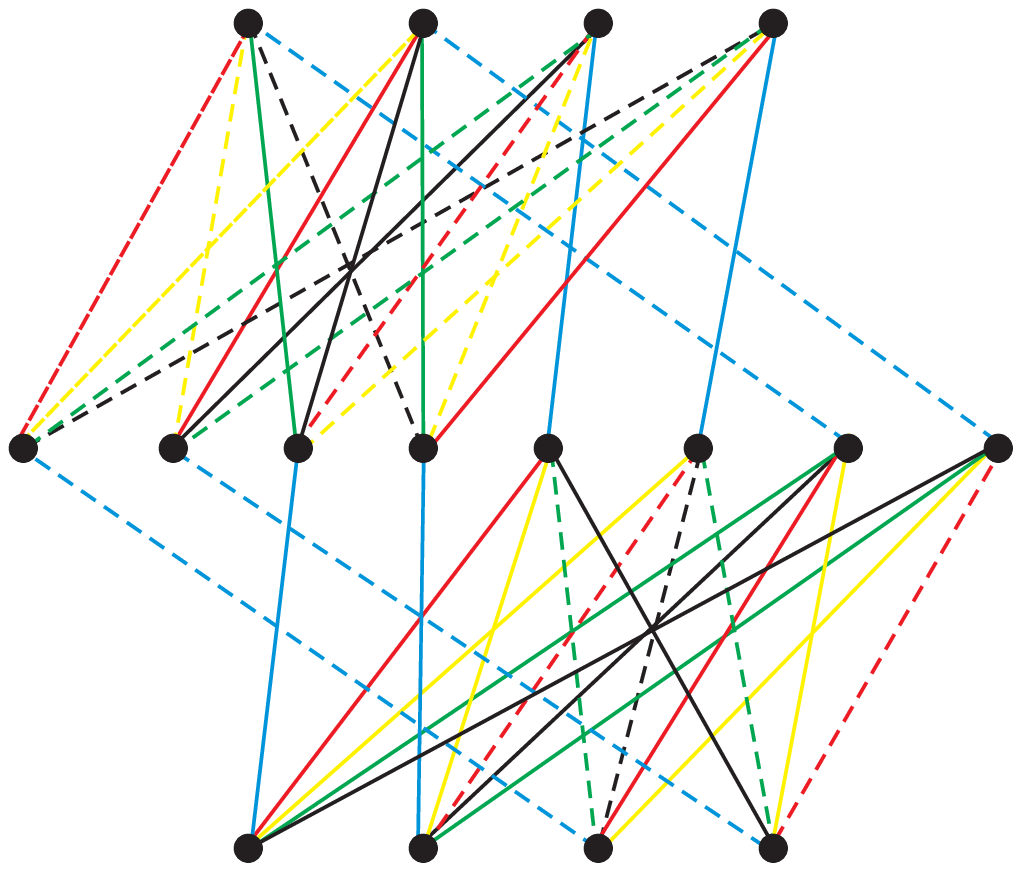} 
\caption{Adinkra of the $N=5$ $(4,8,4)$ multiplet of
$4_4+4_1$ connectivity (type $A$).}
\vspace{1 cm}
\epsfig{file=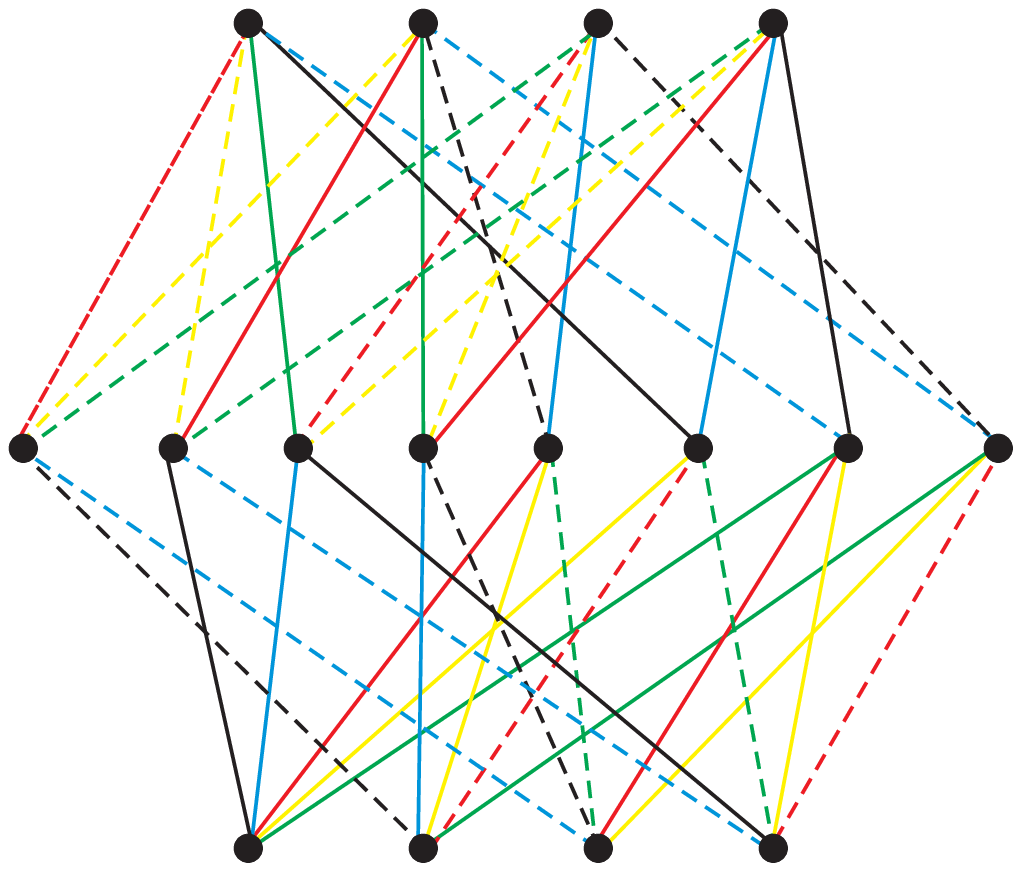} 
\caption{Adinkra of the $N=5$ $(4,8,4)$ multiplet of
$4_3+4_2$ connectivity (type $B$).}
\end{figure}
\newpage

\section{Conclusions}

In this paper we computed the allowed connectivities of the finite linear irreducible representations of
the (\ref{nsusyalgebra}) supersymmetry algebra. For length-$3$ irreps the connectivity is encoded in the
$\psi_g$ symbol (\ref{psig}) which specifies how the fields in an irrep are linked together by supersymmetry
transformations. For $N\leq 8$ we classified which irreps with the same fields content admit different connectivities
(they only exist for $N=5,6$). 
As a corollary, we classified the irreps with inequivalent ``sources and targets".\par
After the first version of this paper appeared, a revision of \cite{dfghil2} was produced.
It was pointed out the existence of an extra irrep, missed in our first version, corresponding to the $N=5$ $(4,8,4)_{A'}$ irrep.
The extra cases w.r.t. our previous version of this paper can only appear for self-dual
$(4,8,4$) multiplets. Besides the $N=5$ $(4,8,4)_{A'}$ irrep, we found a second extra case given by the $ N=6$ $(4,8,4)_{A'}$ irrep.\par
Concerning the \cite{dfghil2} reply to our previous comments, we limit ourselves to point out that we defined and introduced the
$\psi_g$ symbol as a {\em quantitative way} of discriminating irreps of inequivalent connectivities. The quantitative discrimination 
explicitly
discussed in \cite{dfghil2}, based on the number of sources and targets of given heigth, only allows to spot the difference between the
inequivalent irreps appearing in the first version of \cite{dfghil2} (and few extra cases). It fails to spot a difference for the
large class of inequivalent irreps here discussed. 
\par
The approach here discussed can be straighforwardly generalized to compute the connectivities of the $N\geq 9$ irreps
of \cite{krt}.
Concerning physical applications, irreps were classified according to their fields content in \cite{krt}. The differences in
fields content have obvious physical meanings (as already recalled, irreps with different fields content produce, e.g., one-dimensional
supersymmetric sigma models which are embedded in target manifolds of different dimensionality, see \cite{abc}). In order
to understand the physical implications of the irreps with same fields content but different connectivity, it would be quite important to construct off-shell invariant actions for such irreps. As far as we know, the construction of such off-shell invariant actions has not been accomplished yet. 
For $N=8$ a large class of off-shell invariant actions, for each given irrep, has been constructed in \cite{abc}. The list in
\cite{abc} is not exhaustive (see, e.g., \cite{krt}, where an extra off-shell invariant action was produced). It is possible,
but unlikely, that the problem of constructing off-shell invariant actions for multiplets with different connectivities could
be solved with the \cite{abc} formalism of constrained superfields (since we are dealing with $N>4$ systems). It is unclear in fact how to constrain the superfields in the cases under consideration. On the other hand, the linear supersymmetry transformations of the irreps are already given.
It therefore looks promising to use the ``linear" approach developed in \cite{krt}. 
We are planning to address this problem in the  future.
Another issue deserving investigation concerns the puzzling similarities shared by both linear and non-linear representations of
the (\ref{nsusyalgebra}) supersymmetry algebra, see e.g. \cite{bk} for a recent discussion. 
One of the main motivations of the present work concerns the understanding of the features of the large-$N$ supersymmetric quantum mechanical
systems, due to their implications in the formulation of the $M$-theory, see the considerations in \cite{glp} and \cite{top}.
The dimensional reduction of the $11$-dimensional maximal supergravity (thought as the low-energy limit of the $M$-theory)
produces an $N=32$ supersymmetric one-dimensional quantum mechanical system.
\newpage
{\Large{\bf Comments on hep-th/0611060v2}}\\~\\
The wise Reader is warmly invited to skip this part, which adds nothing to the paper and is only intended to reply to
some statements contained in hep-th/0611060v2.\par
The authors of hep-th/0611060v2 stated:\\
1) that in the first version of this paper we misquoted Theorem 4.1 in Ref. [5] of their paper.\par
We point out that\\
{\em a}) We cannot have misquoted Ref. [5] because {\em we did not cite} Ref. [5] (math-ph/0512016).\\
{\em b}) We {\em correctly} quoted the statements made in hep-th/0611060 about the application of Theorem 4.1.\par
Indeed, in  hep-th/0611060, page 4, it is written (let's call it {\em Statement A}):\\
``Theorem 4.1 and Corollary 4.2 from Ref. [5] ensure that every Adinkra is {\bf uniquely specified},
respectively, either by its {\em set of targets} or by its {\em set of sources} and the height assignment of these."
(boldface ours).\\
In page 2 it is explained what sources and targets are:\\
``a node is a {\em source} if no lower node connects to it, and a {\em target} if no higher node connects to it."
\\
Nodes are defined a little before, in the same page 2:\\
``Adinkras represent component bosons in a supermultiplet as white nodes, and fermions as black nodes. A white and a black
node are connected by an edge, ...".\par
The claim in version 2, footnote (5), that the set of sources \\
``Being a {\em subgraph} of the Adinkra, they are specified by their connection to the rest of the Adinkra"\\
is incorrect. Sources or targets are vertices (a set of sources is a set of points). A graph or a subgraph contains points
{\em and edges}. 
 \par
 The missed information in hep-th/0611060 is that the topology of the Adinkra is supposed to be given. In math-ph/0512016
 it is written at page 12: ``{\bf Theorem 4.1} {\em Suppose we are given (1) a topology of an Adinkra (that is, a graph
 that could be the underlying graph of an Adinkra), ...}". \par
 In hep-th/0611060v1, two pairs of examples are discussed s.t. {\em Statement A} correctly applies: the inequivalence in these
 pairs of examples is spotted by the set of sources and targets of given heigth
 (expressed as $(s_1,s_2,s_3)$, $(t_1,t_2,t_3)$). No further information is required.
 In our paper we produced a class of examples s.t. {\em Statement A} no longer applies and further
 information is required. For these cases we introduced and defined the $\psi_g$ (or, equivalently, the $~_x\psi$) symbol to spot the difference
 between inequivalent irreps. In hep-th/0611060v1 no other examples, besides the two pairs, were given and no discussion was made on how to spot the 
 differences for the more general class of cases we produced.
 \\ 
 When discussing our examples in hep-th/0611060v2, the authors are forced to rephrase, in words, the information contained in
 our defined $\psi_g$ and $~_x\psi$ symbols. At page 8 we read ``the distinction is clearly displayed not by their number, but by the
 different connectivity to the rest: In the left-hand side Adinkra, the four source-bosons connect to four of the fermions by a single edge, and by four edges to the other four fermions; not so in the right-hand side one. No field redefinition can erase this {\em topological} distinction." 
To spot the difference between the two Adinkras here the authors employ, without mentioning it, the $~_x\psi$ symbol defined in
this paper (in formula (6) of the previous version, formula (7) of the present version). \par
2) On page 3, in footnote (2) (``We find it hard to pinpoint {\em what} Ref. [12] in fact does claim: ...") and footnote (3)
(``Ref [12] in fact lists {\em two} $N=5$ irreps ...")
the hep-th/0611060 authors act pretending to ignore the second letter sent to them by us and M. Rojas 
(the authors of Ref. [12], namely JHEP {\bf 0603} (2006) 098)
on Dec. 13, 2006, which answered the questions raised. 
For completeness, the letter, e-mailed to all six authors of hep-th/0611060, is reported below.\par
{\em ``Dear colleagues,\\
 
with respect to your reply,
 we summarize our considerations in the following list of statements.
\par
1) In our KRT paper we classified the irreps of the 1D N-susy algebra
  according to the number $n_i$ of fields of dimension $d_i=d_0+i/2$
     entering an irrep.
   The complete list of $(n_1,n_2,...,n_l)$ symbols (from now on the
  ``field content of the irreps"), explicitly presented
   up to $N<= 10$ (formulas 4.10 for $N<=8$, B.3 for $N=9$, B.7 and B.11 for
   $N=10$ and $l>4$, while the $l=2,3$ cases were known from PT) was given.
   $(n_1,...,n_l)$ belongs to the list if and only
if there exists at least one N-irrep with the given field content.
\par
  This is a complete and, as far as we know, correct result
   (no counterexample so far has been found).
\par 
   One valid counterexample would amount to
\par 
   a) find a $(n_1,...,n_l)$ field content incorrectly inserted
      (for some given N) in the list or
 \par
  b) find a $(n_1,..., n_l)$ field content incorrectly not inserted
     (for some given N) in the list.
\par 
 Your two $N=6$ $(6,8,2)$ examples do not qualify to counterexample of the
   above statements ($(6,8,2)$ is a valid field content of $N=6$, already
   present in PT).
\par 
 2) We acknowledge a somewhat confused presentation in KRT of the
   $N=3,5$ mod $8$ irreps, due to the ``double oxidation" of $N=3,5$.
  Unlike the other values of N, the $N=3$ root (length-2) irrep can be
   oxidized to either the $N=(4,0)$ susy or the $N=(3,3)$ pseudosusy. On the
   other hand, the reduction to $N=3$ of both $N=(3,3)$ and $N=(4,0)$ produces one
   and the same irrep (Similarly for $N=5$). This last statement was not clear
   to us when writing KRT. It is clear now
   (it is a consequence of the length-2 $\leftrightarrow$ Clifford algebra 
 correspondence).
\par 
3) Irreps and their equivalence classes: KRT was written for a physical, not
   a mathematical journal. It contains the classification at point 1) as 
 well
   as other results (invariant actions, etc.) relevant for physicists. Less
   important (in a physicist's perspective) issues were skipped. This 
 includes
   the straightforward and somehow formal definition of the class of
   transformations acting on irreps defining an equivalence relation s.t. 
 any
   two irreps with the same field content fall into the same class of
   equivalence (it will be soon presented elsewhere).
   As stated in our previous mail, we have always been aware of the 
 possibility
   that a refinement of our classification could be introduced by choosing
   another group of transformations defining another equivalence relation.
\par 
  Your results in no way contradict our findings. They complement it.
   You do not produce any irrep of, let's say, $N=6$ with, let's say, 
 $(3,7,5,1)$
   field content.
\par 
  In KRT we produced a valid and complete classification of the field
   contents of the irreps. Using a valid analogy (simple Lie algebras over
   C are classified by Dynkin's diagrams, while simple Lie algebras over R
   are obtained by the real forms) we can say that our KRT classification
  corresponds to the ``Dynkin diagrams" of the irreps. Your two (according 
 to
   your definitions) inequivalent $N=6$ $(6,8,2)$ irreps can be regarded as two
   ``real forms" of our unique ``Dynkin diagram".
\par 
 Concerning your reply, you do not formulate any real objection to our
 classification at point 1). You limit to object to our presentation of it
 as a ``complete classification of the irreps".  The fact that the equivalence
relation was not explicitly presented in KRT (for the reasons mentioned at
point 3) does not however authorize you to assume that this equivalence
relation does not exist and/or that your definition of the equivalence
relation is the only acceptable.
\par 
Considering the other points raised in your reply (the status of the $N=5$
 reduced from $N=8$) they can be easily answered. A detailed clarification of
  these issues will be soon produced in a forthcoming paper.
\\ 

 Sincerely Yours,
\\ 
 Zhanna Kuznetsova, Moises Rojas and Francesco Toppan
\\ 
E-mails: zhanna@cbpf.br\\
             mrojas@cbpf.br\\
             toppan@cbpf.br\\
Notes:
\\ 
KRT: ref. Kuznetsova-Rojas-Toppan, JHEP 0603 (2006) 098.\\
PT:  ref. Pashnev-Toppan, JMP 42 (2001) 5257.
"}\\
As promised in the letter, these issues were further clarified by one of us (F.T.)
in hep-th/0612276. 

\newpage
\renewcommand{\theequation}{A.\arabic{equation}}
\setcounter{equation}{0}
\par{~}\\
{\Large{\bf Appendix }}\\{~}\\

We present here for completeness the set (unique up to similarity transformations and an overall sign flipping)
of the seven $8\times 8$ gamma matrices $\gamma_i$ which generate the $Cl(0,7)$ Clifford algebra. The seven
gamma matrices, together with the $8$-dimensional identity ${\bf 1}_8$, are used in the construction of the
$N=5,6,7,8$ supersymmetry irreps, as explained in the main text.  
\begin{eqnarray}\label{gammas07}
{\gamma}_1 = \left(
\begin{array}{cccccccc}
0&0&0&1&0&0&0&0\\
0&0&-1&0&0&0&0&0\\
0&1&0&0&0&0&0&0\\
-1&0&0&0&0&0&0&0\\
0&0&0&0&0&0&0&1\\
0&0&0&0&0&0&-1&0\\
0&0&0&0&0&1&0&0\\
0&0&0&0&-1&0&0&0
\end{array}
\right)&&
{\gamma}_2 = \left(
\begin{array}{cccccccc}
0&1&0&0&0&0&0&0\\
-1&0&0&0&0&0&0&0\\
0&0&0&-1&0&0&0&0\\
0&0&1&0&0&0&0&0\\
0&0&0&0&0&1&0&0\\
0&0&0&0&-1&0&0&0\\
0&0&0&0&0&0&0&-1\\
0&0&0&0&0&0&1&0
\end{array}
\right)\nonumber\\
{\gamma}_3 = \left(
\begin{array}{cccccccc}
0&0&0&0&0&0&1&0\\
0&0&0&0&0&0&0&1\\
0&0&0&0&-1&0&0&0\\
0&0&0&0&0&-1&0&0\\
0&0&1&0&0&0&0&0\\
0&0&0&1&0&0&0&0\\
-1&0&0&0&0&0&0&0\\
0&-1&0&0&0&0&0&0
\end{array}
\right)&&
{\gamma}_4 = \left(
\begin{array}{cccccccc}
0&0&1&0&0&0&0&0\\
0&0&0&1&0&0&0&0\\
-1&0&0&0&0&0&0&0\\
0&-1&0&0&0&0&0&0\\
0&0&0&0&0&0&-1&0\\
0&0&0&0&0&0&0&-1\\
0&0&0&0&1&0&0&0\\
0&0&0&0&0&1&0&0
\end{array}
\right)\nonumber\\
{\gamma}_5 = \left(
\begin{array}{cccccccc}
0&0&0&0&0&1&0&0\\
0&0&0&0&1&0&0&0\\
0&0&0&0&0&0&0&1\\
0&0&0&0&0&0&1&0\\
0&-1&0&0&0&0&0&0\\
-1&0&0&0&0&0&0&0\\
0&0&0&-1&0&0&0&0\\
0&0&-1&0&0&0&0&0
\end{array}
\right)&&
{\gamma}_6 = \left(
\begin{array}{cccccccc}
0&0&0&0&1&0&0&0\\
0&0&0&0&0&-1&0&0\\
0&0&0&0&0&0&1&0\\
0&0&0&0&0&0&0&-1\\
-1&0&0&0&0&0&0&0\\
0&1&0&0&0&0&0&0\\
0&0&-1&0&0&0&0&0\\
0&0&0&1&0&0&0&0
\end{array}
\right)\nonumber\\
{\gamma}_7 = \left(
\begin{array}{cccccccc}
0&0&0&0&0&0&0&1\\
0&0&0&0&0&0&-1&0\\
0&0&0&0&0&-1&0&0\\
0&0&0&0&1&0&0&0\\
0&0&0&-1&0&0&0&0\\
0&0&1&0&0&0&0&0\\
0&1&0&0&0&0&0&0\\
-1&0&0&0&0&0&0&0
\end{array}
\right)&&
{\bf 1}_8 = \left(
\begin{array}{cccccccc}
1&0&0&0&0&0&0&0\\
0&1&0&0&0&0&0&0\\
0&0&1&0&0&0&0&0\\
0&0&0&1&0&0&0&0\\
0&0&0&0&1&0&0&0\\
0&0&0&0&0&1&0&0\\
0&0&0&0&0&0&1&0\\
0&0&0&0&0&0&0&1
\end{array}
\right)
\end{eqnarray}

{}~
\\{}~
\par {\large{\bf Acknowledgments}}{} ~\\{}~\par
This work received support from CNPq (Edital Universal 19/2004) and FAPERJ. Z. K. acknowledges
FAPEMIG for financial support and CBPF for hospitality. We are grateful
to Francisca Val\'{e}ria Fortaleza Gomes for the drawings.

\end{document}